\documentclass[english,american,twocolumn,english]{revtex4-2}
\usepackage[T1]{fontenc}
\usepackage[utf8]{luainputenc}
\usepackage{float}
\usepackage{amsmath}
\usepackage{amssymb}
\usepackage{graphicx}

\makeatletter

\newcommand*\LyXZeroWidthSpace{\hspace{0pt}}

\makeatother

\usepackage{babel}
\begin{document}
\title{\selectlanguage{english}%
Chirality and polarization of inertial antiferromagnetic resonances driven by spin-orbit torques}
\date{\selectlanguage{english}%
\today}
\author{\selectlanguage{english}%
Peng-Bin He}
\email{hepengbin@hnu.edu.cn}

\affiliation{\selectlanguage{english}%
School of Physics and Electronics, Hunan University, Changsha 410082, China}
\author{\selectlanguage{english}%
Ri-Xing Wang}
\affiliation{\selectlanguage{english}%
College of Computer and Electrical Engineering, Hunan University of Arts and Science, Changde 415000, China}
\author{\selectlanguage{english}%
Zai-Dong Li}
\affiliation{\selectlanguage{english}%
Tianjin Key Laboratory of Quantum Optics and Intelligent Photonics, School of Science, \\ Tianjin University of Technology, Tianjin 300384, China \\
and School of Mathematics and Physics, Xinjiang Hetian College, Hetian 848000, China}
\author{\selectlanguage{english}%
Mikhail Cherkasskii}
\email{macherkasskii@hotmail.com}

\affiliation{\selectlanguage{english}%
Institute for Theoretical Solid State Physics, RWTH Aachen University, DE-52074 Aachen, Germany}
\begin{abstract}
It is widely accepted that the handedness of a resonant mode is an intrinsic property. We show that, by tailoring the polarization and handedness of alternating spin-orbit torques used as the driving force, the polarization state and handedness of inertial resonant modes in an antiferromagnet (AFM) can be actively controlled. In contrast with ferromagnets, whose resonant-mode polarization is essentially fixed, AFM inertial modes can continuously evolve from elliptic through circular to linear polarization as the driving polarization is varied. We further identify an inertia-dependent critical degree of driving polarization at which the mode becomes linearly polarized while its handedness reverses.
\end{abstract}
\maketitle

\section{introduction}

\label{introduction}

Spin inertia \citep{RMondal} in magnetically ordered materials can be induced by various microscopic mechanisms, as demonstrated in recent studies. First-principles spin dynamics have shown that both damping and inertia emerge consistently from the electronic structure, emphasizing the role of spin-orbit coupling in dissipative and inertial spin dynamics \citep{SBhattacharjee,RMondal_JPCM}. A complementary approach based on spin-bath coupling reveals that a non-Markovian memory kernel yields an effective inertial term in the spin equation of motion \citep{MAQuarenta,JAnders}. Additionally, nonadiabatic interactions with environmental degrees of freedom can produce inertia-like corrections \citep{TKikuchi}. A classical interpretation models the spin as a circular current loop, leading to inertial terms from the underlying mechanical dynamics~\citep{SGiordano,JEWegrowe}.

Spin inertia gives rise to magnetic nutation \citep{MCCiornei,DBottcher,TKikuchi,SVTitov_PRB103,SVTitov}, which manifests as nutation resonance \citep{YLi,KNeeraj,VUnikandanunni,ADe}. Furthermore, inertia modifies the self-oscillatory dynamics of magnetization in both antiferromagnets (AFMs) \citep{PBHe_PRB108,PBHe_PRB110} and ferromagnets (FMs) \citep{RRodriguez}. The characteristic influence of spin inertia on these uniform magnetic dynamics can serve as an additional experimental probe for its detection.

Apart from experimental observations of nutation \citep{YLi,KNeeraj,VUnikandanunni,ADe}, extensive efforts have been invested in the theoretic research of nutational resonance \citep{EOlive_APL,EOlive_JAP,MCherkasskii_PRB102,RMondal_PRB103,RMondal_PRB104,RMondal_PRB104_21,MondalR_JPCM33,SVTitov_JAP,MCherkasskii_PRB106,SGhosh}. It was found that the inertial magnetization dynamics are described by the inertial Landau-Lifshitz-Gilbert (ILLG) equation, which predicts a nutation resonance in addition to conventional FM resonance. Olive et al. \citep{EOlive_APL,EOlive_JAP} numerically demonstrated a secondary high-frequency peak whose amplitude and position depend on the inertial relaxation time, the damping, and the applied field. In subsequent analytical studies shifts of resonant frequencies and linewidths of both precession and nutation have been quantified, revealing a nontrivial dependence on damping and magnetic anisotropy \citep{MCherkasskii_PRB102,MCherkasskii_PRB106,SVTitov_JAP,SGhosh}. Authors of another analytical study have predicted that these effects are enhanced in AFMs due to exchange interaction \citep{RMondal_PRB103}. Remarkably, when driven by a circularly polarized magnetic field, FMs exhibit a single nutational peak and a single precessional peak, whereas AFMs and ferrimagnets present these peaks in pairs. In the latter two systems, inter- and intrasublattice couplings further tune these resonances, indicating distinct mechanisms in multisublattice systems~\citep{RMondal_PRB104,MondalR_JPCM33}. Finally, spin pumping at terahertz nutation frequencies produces a spin current opposite in direction to that generated by precessional modes, providing an unambiguous experimental signature of nutation \citep{RMondal_PRB104_21}.

We note that authors of most studies on inertial resonances have primarily focused the resonant frequency, determined by the intrinsic magnetic properties. However, the polarization of the resonant mode, which depends on both the magnetic material and the external drive, remains comparatively unexplored, despite offering an extra degree of freedom -- in addition to amplitude and phase -- for information encoding \citep{RCheng,TYu,WYu,CJia}. Experimentally, resonance is usually driven by circularly or linearly polarized magnetic fields. Accordingly, the nutational resonances in the current theoretical studies were assumed to be excited by circularly \citep{MCherkasskii_PRB102,RMondal_PRB103,RMondal_PRB104,RMondal_PRB104_21,MondalR_JPCM33,SGhosh} or linearly polarized \citep{EOlive_APL,EOlive_JAP} oscillating fields.

An alternative utilizes current-induced spin-torque resonance, realized via spin-transfer torques in spin-valve structures \citep{AATulapurkar,JCSankey,HXi,JNKupferschmidt,AAKovalev,PBHe} and spin-orbit torques (SOTs) in heavy-metal/FM bilayers \citep{LLiu,KKondou,VSluka,CSun}, both of which are efficient and scalable in spintronic devices. In both schemes an alternating charge current generates an oscillating spin current that exerts a spin torque on the magnetization. Particularly, in SOT-based devices the in-plane current is more easily tuned than the perpendicular current, enabling finer control of the alternating driving torque. In addition, the role of the inertial effects in mode polarization has received limited attention in the literature. Authors of prior studies on spin dynamics driven by SOTs in inertial magnetic materials have implicitly assumed that the handedness and polarization are essentially fixed by the material. A systematic investigation of the polarization of resonant modes and its tuning by driving forces is, to the best of our knowledge, lacking in the literature.

Motivated by the above considerations, we present the investigation of inertial resonant modes in AFMs driven by alternating SOTs. We show that SOTs generated by two orthogonal in-plane currents tune the polarization state (linear, elliptical or circular) and handedness (left or righthanded) of both precessional and nutational AFM resonances. We find that the handedness reverses at critical current ratios that depend explicitly on the inertial relaxation time. These inertia-dependent switching points provide a practical route to determine spin inertia across a broad range of materials. We use chirality and handedness interchangeably, following common usage in the literature.

\section{model}

\label{model}

\begin{figure}[b]
\includegraphics[width=8cm]{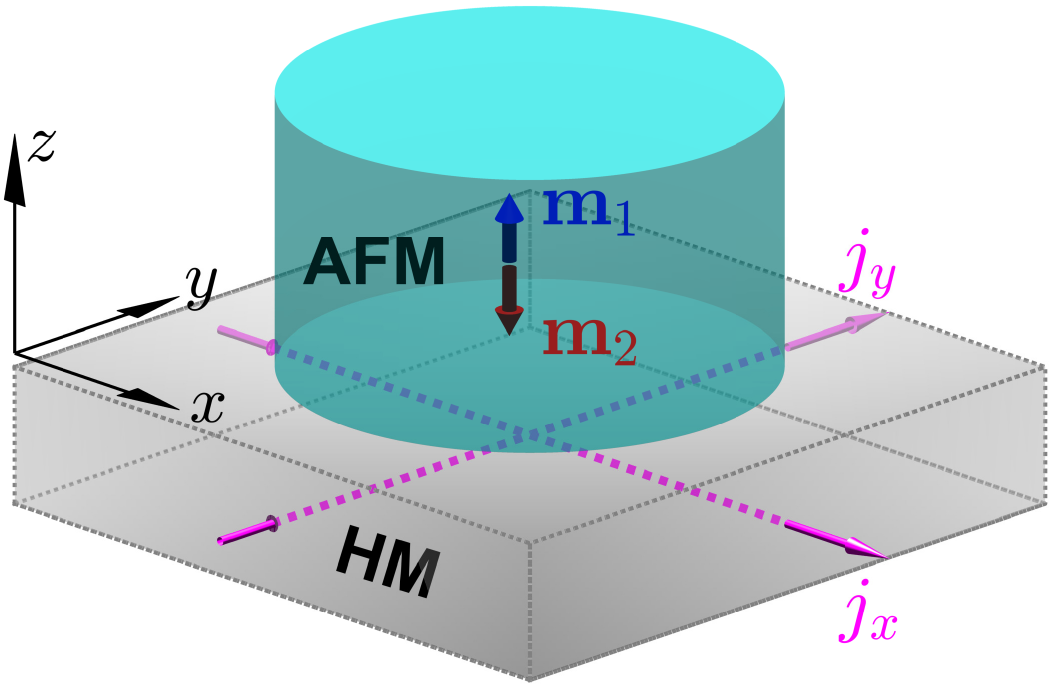}

\caption{(color online). Schematic diagram of model.}
\label{device}
\end{figure}

As illustrated in Fig.~\ref{device}, we study an inertial AFM film attached to a heavy metal (HM) with strong spin-orbit coupling, driven by oscillating in-plane currents. The bipartite AFM
dynamics are captured phenomenologically by two exchange-coupled ILLG equations, which include spin inertia, Gilbert damping, and SOTs. For sublattice $k$, the ILLG equation reads: 
\begin{equation}
\frac{d\mathbf{m}_{k}}{dt}=\mathbf{m}_{k}\times\frac{d\mathcal{E}}{d\mathbf{m}_{k}}+\alpha\mathbf{m}_{k}\times\frac{d\mathbf{m}_{k}}{dt}+\eta\mathbf{m}_{k}\times\frac{d^{2}\mathbf{m}_{k}}{dt^{2}}+\boldsymbol{\tau}_{k},\label{LLG equation}
\end{equation}
where $\mathbf{m}_{k}$ is the unit magnetization vector of sublattice $k$ ($k=1,2$), $\alpha$ is the Gilbert damping constant, and $\eta$ is the inertial relaxation time, typically ranging from femtoseconds to picoseconds, as predicted by \textit{ab initio} calculations \citep{DThonig} and confirmed experimentally \citep{YLi,KNeeraj,VUnikandanunni}.

The magnetic energy, arising from the magnetocrystalline anisotropy and the intersublattice exchange interaction, is expressed in frequency units as 
\begin{equation}
\mathcal{E}=\omega_{E}\mathbf{m}_{1}\cdot\mathbf{m}_{2}+\omega_{K}\sum_{k}\left(\mathbf{m}_{k}\cdot\mathbf{e}_{z}\right)^{2},\label{magnetic energy}
\end{equation}
where $\omega_{E(K)}=\gamma_{0}H_{E(K)}$, with $H_{E}$ and $H_{K}$ being the exchange field and the anisotropy field, respectively. Here, $\gamma_{0}=g\mu_{0}\mu_{B}/\hbar$ is the gyromagnetic ratio with $g$ being the Landé $g$ factor, $\mu_{0}$ the vacuum susceptibility, $\mu_{B}$ the Bohr magneton, and $\hbar$ the reduced Plank constant. The anisotropy easy axis is oriented perpendicular to the film plane.

The last term in Eq. (\ref{LLG equation}) represents the SOTs, originating from spin transfer between the local magnetization and spin currents generated in the adjacent HM layer. Including both dampinglike and fieldlike components \citep{AManchon}, the SOTs are given by 
\begin{equation}
\mathbf{\boldsymbol{\tau}}_{k}=-\rho\left\{ \mathbf{m}_{k}\!\times\!\left[\mathbf{m}_{k}\!\times\!\left(\mathbf{e}_{z}\!\times\!\mathbf{j}_{e}\right)\right]\!+\!\beta\mathbf{m}_{k}\!\times\!\left(\mathbf{e}_{z}\!\times\!\mathbf{j}_{e}\right)\right\} ,\label{SOTs}
\end{equation}
where $\beta$ is the ratio of fieldlike to dampinglike SOT.

The prefactor $\rho=\xi\mu_{B}/(eM_{s}d)$ sets the overall SOT strength. Here $M_{s}$ is the saturation magnetization of sublattice, $d$ denotes the thickness of the AFM layer, and $e$ is the elementary charge. The SOT efficiency is defined as $\xi=T_{\mathrm{int}}\theta_{\mathrm{sh}}$ \citep{LZhu,CFPai}, where $\theta_{\mathrm{sh}}$ is the spin Hall angle \citep{AManchon} and $T_{\mathrm{int}}$ the spin transparency \citep{ZLWang} of the HM/AFM interface. The dampinglike SOT is nonconservative. It represents the antidamping torque that is even under $\mathbf{m}_{k}\to-\mathbf{m}_{k}$. It can be written as a torque from an effective field $\mathbf{H}_{\mathbf{DL}}\propto\mathbf{m}_{k}\times\mathbf{p},$ where $\mathbf{p}\equiv\mathbf{e}_{z}\times\mathbf{j}_{e}$, thereby modifying the effective Gilbert damping in linear response. The fieldlike SOT is precessional conservative torque that is odd under $\mathbf{m}_{k}\to-\mathbf{m}_{k}$. It induces precession about an effective field $\mathbf{H}_{\mathrm{FL}}\propto\mathbf{p},$where $\mathbf{p}$ is the spin-polarization direction set by the spin Hall effect in the HM.

The in-plane electric current density $\mathbf{j}_{e}$ is expressed in components as 
\begin{equation}
\mathbf{j}_{e}=j_{x}\mathbf{e}_{x}+j_{y}\mathbf{e}_{y}.
\end{equation}
To excite stationary modes, in-plane alternating electric currents are applied, as shown in Fig.~\ref{device}. For analytical convenience, the current is expressed in complex form $(j_{x},j_{y})=(J_{x},iJ_{y})e^{-i\omega t}$, where $J_{x(y)}>0$ denote the current amplitudes. In this representation, the phase of $j_{y}$ leads (lags) that of $j_{x}$ by $90^{\circ}$ for positive (negative) $\omega$, corresponding to right-handed (left-handed) polarized SOTs. Taking the real part yields the physical current: $J_{x}\cos\omega t\,\mathbf{e}_{x}+J_{y}\sin\omega t\,\mathbf{e}_{y}$.

Each term in the ILLG equation encodes a distinct \foreignlanguage{american}{physical} mechanism. The precessional term, $\mathbf{m}_{k}\times d\mathcal{E}/d\mathbf{m}_{k}$ drives conservative motion in the effective field $\mathbf{H}_{{\rm eff}}=d\mathcal{E}/d\mathbf{m}_{k}$ set by the magnetic \foreignlanguage{american}{potential energy}. The Gilbert term represents dissipation due to coupling of the magnetic moments to a thermal bath. The spin-inertial term accounts for the dynamics related to the kinetic energy. The spin--orbit terms captures the external torques exerted by the HM layer on the AFM film. Although each contribution is well understood, their competition produces the nontrivial dynamics analyzed in Secs.~III-V.

\selectlanguage{american}%
One of the issues in inertial spin dynamics is the applicable time scale, set by the inertial relaxation time. Reported values span orders of magnitude: estimates near $1$~fs were proposed in Ref.~\citep{SBhattacharjee} and inferred from FM-resonance precession data in Ref.~\citep{YLi}. First-principles calculations \citep{DThonig} gave values $\eta\approx1-100$~fs. \emph{Ab initio} simulations \citep{Bouaziz2019} of the dynamical magnetic susceptibility indicated $\eta\sim10-100$~fs. Time-resolved magneto-optical detection of resonant nutation \citep{KNeeraj} yielded $\eta\sim100$~fs. Measurements on cobalt films \citep{VUnikandanunni} reported $\eta/\alpha\approx750$~fs. Given the wide range of reported values for the inertial relaxation time, we present results as a function of $\eta.$

\selectlanguage{english}%

\section{resonant spectrum}

\label{resonant spectrum}

Due to the uniaxial anisotropy and intersublattice exchange {[}Eq. (\ref{magnetic energy}){]}, the static equilibrium magnetizations of the two sublattices align along the $\pm z$ directions, respectively. For small deviations from equilibrium, the $z$-components of the magnetizations remain unchanged to first order, such that $m_{1}^{z}\approx1$ and $m_{2}^{z}\approx-1$. To describe the magnetic response of the AFM in the linear regime, we adopt the ansatz $\mathbf{m}_{k}=(-1)^{k+1}\mathbf{e}_{z}+m_{k}^{x}(t)\mathbf{e}_{x}+m_{k}^{y}(t)\mathbf{e}_{y}$, where $m_{k}^{x}(t)$, $m_{k}^{y}(t)$ are complex-valued functions encoding the amplitude and phase of the response to SOTs. Inserting the ansatz into Eq. (\ref{LLG equation}), assuming a harmonic time dependence of $m_{k}^{x}(t)$, $m_{k}^{y}(t)$ (i.e., replacing $d/dt$ by $-i\omega$) and keeping only terms linear in $m_{k}^{x}(t)$, $m_{k}^{y}(t)$, $j_{x}$ and $j_{y}$, one obtains a four-dimensional linear inhomogeneous system of ordinary differential equations that describes damped magnetic oscillations driven by periodic forces. As detailed in Appendix \ref{app A}, solving this system yields the steady-state linear response modes: 
\begin{equation}
\left[\begin{array}{c}
m_{1}^{x}(t)\\
m_{1}^{y}(t)\\
m_{2}^{x}(t)\\
m_{2}^{y}(t)
\end{array}\right]=\frac{\rho}{\Delta}\left(\begin{array}{cc}
\chi_{1} & \chi_{2}\\
-\chi_{2} & \chi_{1}\\
-\chi_{1} & \chi_{2}\\
-\chi_{2} & -\chi_{1}
\end{array}\right)\left[\begin{array}{c}
j_{x}(t)\\
j_{y}(t)
\end{array}\right].\label{linear modes}
\end{equation}
where $\chi_{1}=(\eta\omega^{2}-2\omega_{E}-\omega_{K})+i(\alpha-\beta)\omega$, $\chi_{2}=\beta(\eta\omega^{2}-\omega_{K})+i(1+\alpha\beta)\omega$, and 
\begin{equation}
\Delta=\left(\Omega_{1}-i\alpha\omega\right)\left(\Omega_{2}-i\alpha\omega\right)-\omega^{2},
\end{equation}
with 
\begin{eqnarray}
\Omega_{1} & = & 2\omega_{E}+\omega_{K}-\eta\omega^{2},\label{Omega_1}\\
\Omega_{2} & = & \omega_{K}-\eta\omega^{2}.\label{Omega_2}
\end{eqnarray}
In Eq. (\ref{linear modes}), a susceptibility tensor is defined to characterize the linear response to alternating SOTs, in analogy with the Polder susceptibility tensor \citep{DDStancil} describing the linear response to an alternating magnetic field.

To analyze the properties of the steady-state modes, it is convenient to take the real part of Eq. (\ref{linear modes}). The resulting harmonic oscillations of the magnetization driven by alternating SOTs with frequency $\omega$ are given by 
\begin{equation}
m_{k}^{x(y)}=\mathcal{A}_{k}^{x(y)}\cos\left(\omega t+\phi_{k}^{x(y)}\right).\label{spin waves}
\end{equation}
The polarization degree of the linear modes is set by the ratio $\mathcal{A}_{k}^{y}/\mathcal{A}_{k}^{x}$, while the phase difference between $m_{k}^{x}$ and $m_{k}^{y}$ determines the handedness of $\mathbf{m}_{k}$. In Eq. (\ref{spin waves}) the amplitudes are 
\begin{eqnarray}
\mathcal{A}_{1,2}^{x(y)}=\rho\sqrt{\frac{P_{1}J_{x(y)}^{2}+P_{2}J_{y(x)}^{2}\pm P_{3}J_{x}J_{y}}{\widetilde{\Delta}}},\label{amplitude}
\end{eqnarray}
where upper and lower signs represent sublattices $1$ and $2$, respectively. This sign convention is maintained throughout unless stated otherwise. To simplify the expressions, we introduce $P_{1}=(\alpha-\beta)^{2}\omega^{2}+\Omega_{1}^{2}$, $P_{2}=(1+\alpha\beta)^{2}\omega^{2}+\beta^{2}\Omega_{2}^{2}$, $P_{3}=2\omega[2(1+\alpha\beta)\omega_{E}+(1+\beta^{2})\Omega_{2}]$, and 
\begin{equation}
\widetilde{\Delta}=\left[\left(1\!-\!\alpha^{2}\right)\omega^{2}\!-\!\Omega_{1}\Omega_{2}\right]^{2}\!+\!4\alpha^{2}\omega^{2}\left(\omega^{2}\!+\!\omega_{E}^{2}\right).\label{Delta}
\end{equation}

\begin{figure}[htbp]
\includegraphics{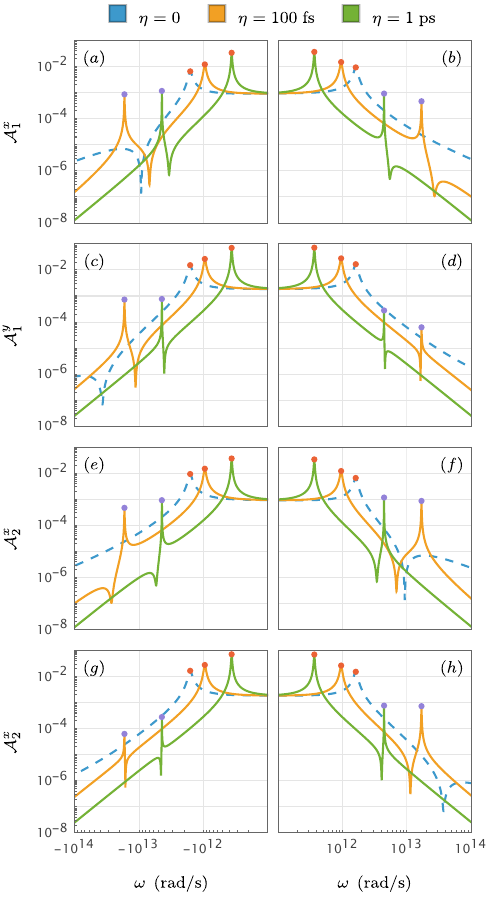}\caption{(color online). AFM spectra without spin inertia exhibit the single precessional resonance (dashed lines, $\eta=0$). Finite spin inertia $\eta>0$ shifts the precessional AFM resonance and induces the additional nutational resonances (solid lines). Peaks labeled with red (purple) points correspond to precessional (nutational) resonances. Both axes are shown on a logarithmic scale; low-frequency parts are omitted for clarity. Curves are evaluated from Eq. (\ref{amplitude}) using $\omega_{E}=9.25$ THz, $\omega_{K}=0.14$ THz, $\alpha=0.01$, $\beta=0.02$, $J_{x}=1$ GA$/$m$^{2}$, $J_{y}=2$ GA$/$m$^{2}$, and $\rho\approx0.13$ Hz$/$(A$/$m$^{2}$). The values of $\omega_{E}$, $\omega_{K}$, and $\rho$ are derived from the magnetic parameters of MnF$_{2}$ \citep{PVaidya} for a film thickness $d=3$nm and SOT efficiency $\xi=0.32$.}
\label{spectrum_all}
\end{figure}

The phases in Eq. (\ref{spin waves}) are written as 
\begin{eqnarray}
\phi_{1,2}^{x} & = & \tan^{-1}\left(\pm Q_{1}J_{x}\!+\!Q_{2}J_{y},\pm Q_{3}J_{x}\!+\!Q_{4}J_{y}\right),\label{phase of mx}\\
\phi_{1,2}^{y} & = & \tan^{-1}\left(Q_{4}J_{x}\!\pm\!Q_{3}J_{y},-Q_{2}J_{x}\!\mp\!Q_{1}J_{y}\right),\label{phase of my}
\end{eqnarray}
where $\tan^{-1}\left(x,y\right)$ denotes the two-argument arctangent that returns the angle of the vector $(x,y)$, correctly resolving the quadrant \citep{arctan}. In Eqs. (\ref{phase of mx}) and (\ref{phase of my}), the coefficients before $J_{x,y}$, relying on $\omega$, are given by $Q_{1}=(1+\alpha\beta)\Omega_{1}\omega^{2}-[\alpha(\alpha-\beta)\omega^{2}+\Omega_{1}^{2}]\Omega_{2}$, $Q_{2}=\omega\{(1+\alpha^{2})(1+\alpha\beta)\omega^{2}+[\alpha\beta\Omega_{2}-\Omega_{1}]\Omega_{2}\}$, $Q_{3}=\omega\{(1+\alpha^{2})(\alpha-\beta)\omega^{2}+[\alpha\Omega_{1}+\beta\Omega_{2}]\Omega_{1}\}$, and $Q_{4}=(\alpha-\beta)\Omega_{2}\omega^{2}+[\alpha(1+\alpha\beta)\omega^{2}+\beta\Omega_{2}^{2}]\Omega_{1}$. Eqs. (\ref{phase of mx}) and (\ref{phase of my}) show that the phase lag of the stationary forced oscillations relative to the driving SOTs depends on the amplitude ratio $J_{x}/J_{y}$ and on the driving frequency $\omega$.

The handedness of the linear modes is determined by the phase difference between the $x$- and $y$-components of $\mathbf{m}_{1,2}$. From Eqs. (\ref{phase of mx}) and (\ref{phase of my}) one obtains the difference 
\begin{equation}
\Delta\phi_{1,2}\equiv\phi_{1,2}^{x}-\phi_{1,2}^{y}=\tan^{-1}\left(X_{1,2},Y_{1,2}\right),\label{phase dfferences}
\end{equation}
where $X_{1,2}=\pm(Q_{1}Q_{4}-Q_{2}Q_{3})(J_{x}^{2}-J_{y}^{2})$, and $Y_{1,2}=\pm(Q_{1}Q_{2}+Q_{3}Q_{4})(J_{x}^{2}+J_{y}^{2})+(Q_{1}^{2}+Q_{2}^{2}+Q_{3}^{2}+Q_{4}^{2})J_{x}J_{y}$. Note that when flipping the sign of $\omega$, $Q_{1}$ and $Q_{4}$ remain unchanged and $Q_{2}$ and $Q_{3}$ change the sign. Consequently, $X_{1,2}$ do not change and $Y_{1}$ ($Y_{2}$) becomes $Y_{2}$ ($Y_{1}$).

The system shows two SOT--driven resonances: a low-frequency precession and a high-frequency nutation created by spin inertia (Fig.~\ref{spectrum_all}). Inertia enters via the $-\eta\omega^{2}$ terms in $\Omega_{1,2}$, shifting the spectrum and producing the nutation peak. One can see that the precessional resonances are stronger (higher peaks) and occur at lower frequencies than nutational ones. However, both resonances lie in the terahertz range, due to the exchange enhancement characteristic of AFMs. Figure \ref{spectrum_all} shows the precessional and nutational spectra for different inertial relaxation times $\eta$, considering the general case $J_{x}\neq J_{y}$. The left (right) column displays spectra at negative (positive) frequencies. It is important to note that panels (a) and (b), (c) and (d), (e) and (f), and (g) and (h) reveal an asymmetry of the spectra with respect to frequency inversion for a given sublattice. Additionally, comparing panels (a) and (f), (b) and (e), (c) and (h), and (d) and (g), one observes that the positive-frequency spectrum of $m_{1}^{x,y}$ mirrors the negative-frequency spectrum of $m_{2}^{x,y}$ and vice versa. This can be understood from symmetry arguments: Exchanging the sublattices ($\mathcal{A}_{1}^{x,y}\leftrightarrow\mathcal{A}_{2}^{x,y}$) is equivalent to time reversal ($\omega\leftrightarrow-\omega$). Therefore, the spectra are symmetric under the combined transformation $1\leftrightarrow2$ and $\omega\leftrightarrow-\omega$.

\begin{figure*}[t]
\selectlanguage{american}%
\includegraphics{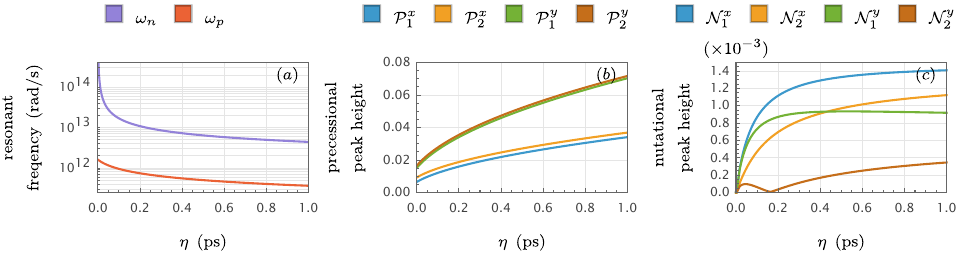}

\selectlanguage{english}%
\caption{(color online). (a) Resonant frequencies as a function of $\eta$. Curves are computed from Eq. (\ref{resonant frequency}). A logarithmic scale is used on the frequency axis, and low-frequency regions are omitted for clarity. (b) and (c) show the dependence of peak heights on $\eta$ for precession and nutation, respectively. Curves are computed from Eqs. (\ref{peak heights of precession}) and (\ref{peak heights of nutation}). The parameters used are the same as in Fig. \ref{spectrum_all}.}
\label{resonant_frequency_and_height_vs_eta}
\end{figure*}

\section{Resonant peak and frequency}

\label{resonant frequency and peak height}

From Eqs. (\ref{amplitude}) and (\ref{Delta}), and noting that $\alpha^{2}\ll1$, the amplitudes of the linear modes increases at resonance, which occurs when 
\begin{equation}
\omega^{2}=\Omega_{1}\Omega_{2},\label{resonant equation}
\end{equation}
where $\Omega_{1,2}$ are defined in Eqs. (\ref{Omega_1}) and~(\ref{Omega_2}). Solving Eq. (\ref{resonant equation}) yields the resonant frequencies, which appear in pairs $\pm\omega$. In the absence of spin inertia ($\eta=0$), the resonance condition gives $\omega=\pm\sqrt{\omega_{K}(2\omega_{E}+\omega_{K})}$. In the presence of inertia, the nutational frequency $\omega_{n}$ and the precessional frequency $\omega_{p}$ are given by 
\begin{eqnarray}
\omega_{n,p}=\sqrt{\left(\frac{\kappa_{n,p}}{2\eta}\right)^{2}-\omega_{E}^{2}},\label{resonant frequency}
\end{eqnarray}
where 
\begin{equation}
\kappa_{n,p}=\sqrt{\left(1+2\eta\omega_{E}\right)^{2}+4\eta\omega_{K}}\pm1,\label{kappa}
\end{equation}
with the upper (lower) sign corresponding to nutation (precession). The resonant frequencies exhibit strong dependence on the inertial relaxation time $\eta$, as shown in Fig. \ref{resonant_frequency_and_height_vs_eta}a.

Since the exchange interaction dominates the dynamics in AFMs, the precessional resonance is enhanced into the terahertz regime. At the same time, the nutational frequency is significantly higher than in FMs. Moreover, spin inertia induces a pronounced redshift in the resonant frequencies of both nutation and precession, as also evident in the spectra shown in Fig. \ref{spectrum_all}.

In general, resonant peak heights quantify the energy stored in the forced oscillations. As illustrated in Fig. \ref{spectrum_all}, these heights depend on the inertial relaxation time~$\eta$ and differ for positive and negative frequencies. Therefore, it is necessary to analyze them in detail. Substituting the resonant frequencies (\ref{resonant frequency}) into Eqs. (\ref{amplitude}) and (\ref{Delta}) yields explicit expressions for the peak heights, which are different for each sublattice and each component of magnetization. For the $\omega_{n}$-branch resonance, the peak heights of $x(y)$ components of sublattices $1$ and $2$ are as follows: 
\begin{eqnarray}
\mathcal{N}_{1,2}^{x(y)} & = & \rho\frac{\eta}{\sqrt{2}\alpha\kappa_{n}}\sqrt{\frac{\kappa_{n}\!-\!2\eta\omega_{E}}{2\eta\left(\omega_{E}\!+\!\omega_{K}\right)\!+\!\kappa_{n}}}\nonumber \\
 &  & \times \left\vert \!\sqrt{\kappa_{n}\!-\!2\eta\omega_{E}}J_{x(y)}\!\mp\!\sqrt{\kappa_{n}\!+\!2\eta\omega_{E}}J_{y(x)}\!\right\vert. \label{peak heights of nutation}
\end{eqnarray}
For the $\omega_{p}$-branch resonance, the heights of $x(y)$ components of sublattices $1$ and $2$ are given by 
\begin{eqnarray}
\mathcal{P}_{1,2}^{x(y)} & = & \rho\frac{\eta}{\sqrt{2}\alpha\kappa_{p}}\sqrt{\frac{\kappa_{p}\!+\!2\eta\omega_{E}}{2\eta\left(\omega_{E}\!+\!\omega_{K}\right)\!-\!\kappa_{p}}}\nonumber \\
 &  & \times \left\vert \!\sqrt{\kappa_{p}\!-\!2\eta\omega_{E}}J_{x(y)}\!\pm\!\sqrt{\kappa_{p}\!+\!2\eta\omega_{E}}J_{y(x)}\!\right\vert. \label{peak heights of precession}
\end{eqnarray}
The peak heights of the negative ($-\omega_{n}$ and $-\omega_{p}$) branches are obtained by interchanging sublattices $1$ and $2$ in Eqs. (\ref{peak heights of nutation}) and (\ref{peak heights of precession}), respectively. Throughout the derivation we assume $\alpha^{2}\!\ll\!1$, $\alpha\beta\!\ll\!1$, and $\beta^{2}\!\ll\!1$.

Equations (\ref{peak heights of nutation}) and (\ref{peak heights of precession}) reveal several notable features. (i) The peak heights are highly sensitive to the relative magnitudes of $J_{x}$ and $J_{y}$. For the generic case $J_{x}\neq J_{y}$, as shown in Figs.~\ref{resonant_frequency_and_height_vs_eta}b and~\ref{resonant_frequency_and_height_vs_eta}c, the $x$ and $y$ components exhibit unequal peak heights, indicating elliptical polarization of the resonant modes. Spin inertia further enhances the ellipticity. (ii) Only when $J_{x}=J_{y}$ do the $x$ and $y$ components exhibit equal peak heights, corresponding to circularly polarized modes. In this case, all four resonances at $\pm\omega_{n}$ and $\pm\omega_{p}$ exist. This contrasts with the FM case, where only two modes ($-\omega_{n}$ and $\omega_{p}$) appear, as seen in Figs.~\ref{spectrum_all_FM}e and~\ref{spectrum_all_FM}f. (iii) The nutational peaks are generally much smaller than the precessional ones, as evident from comparison between Figs. \ref{resonant_frequency_and_height_vs_eta}b and \ref{resonant_frequency_and_height_vs_eta}c. In the limit $\eta=0$, one has $\kappa_{n}=2$ and thus $\mathcal{N}_{1,2}^{x}=\mathcal{N}_{1,2}^{y}=0$, indicating the absence of nutation. Meanwhile, $\kappa_{p}=0$ for $\eta=0$. Then taking the limit $\eta\rightarrow0$ in Eq. (\ref{peak heights of precession}), the precessional peak heights approach 
\begin{equation}
\frac{\rho}{2\alpha}\frac{\left\vert \sqrt{2\omega_{E}+\omega_{K}}J_{x(y)}\pm\sqrt{\omega_{K}}J_{y(x)}\right\vert }{\left(\omega_{E}+\omega_{K}\right)\sqrt{\omega_{K}}},
\end{equation}
which corresponds to the resonant peak heights of AFM sublattices in the absence of spin inertia. (iv) For specific ratios of $J_{x}$ and $J_{y}$, certain nutational peak vanishes. For example, when $J_{y}/J_{x}=\sqrt{(\kappa_{n}+2\eta\omega_{E})/(\kappa_{n}-2\eta\omega_{E})}$, one finds $\mathcal{N}_{1}^{y}=0$, as indicated by the brown curve in Fig. \ref{resonant_frequency_and_height_vs_eta}c. Complete suppression of this nutational peak enables polarization tuning from circular to linear.

\section{Analyses of polarization and handedness}

\label{polarization and handedness}

To characterize the resonance polarization we employ two quantities: the ellipticity, which measures the degree of polarization, and the handedness, extracted from the phase difference between $m_{x}$ and $m_{y}$. The ellipticity is defined as the aspect ratio $m_{y}/m_{x}$: It equals $1$ for circular polarization and tends to $0$ or $\infty$ for linear polarization; intermediate values correspond to elliptical polarization. For the nutational modes we introduce ellipticity as $e_{1,2}^{n}=\mathcal{N}_{1,2}^{y}/\mathcal{N}_{1,2}^{x}$, and for the precessional modes $e_{1,2}^{p}=\mathcal{P}_{1,2}^{y}/\mathcal{P}_{1,2}^{x}$, where the subscripts $1,2$ label the two sublattices. Using Eqs. (\ref{peak heights of nutation}) and (\ref{peak heights of precession}) we obtain 
\begin{eqnarray}
e_{1,2}^{n} & = & \left\vert \frac{r_{n}J_{x}\mp J_{y}}{J_{x}\mp r_{n}J_{y}}\right\vert ,\label{en}\\
e_{1,2}^{p} & = & \left\vert \frac{r_{p}J_{x}\pm J_{y}}{J_{x}\pm r_{p}J_{y}}\right\vert ,\label{ep}
\end{eqnarray}
where 
\begin{equation}
r_{n,p}(\eta)=\sqrt{\frac{\sqrt{\left(1+2\eta\omega_{E}\right)^{2}+4\eta\omega_{K}}\pm1+2\eta\omega_{E}}{\sqrt{\left(1+2\eta\omega_{E}\right)^{2}+4\eta\omega_{K}}\pm1-2\eta\omega_{E}}}.\label{rnp}
\end{equation}
In the limit $\eta\to0$ one recovers $r_{p}=\sqrt{(2\omega_{E}+\omega_{K})/\omega_{K}}$ and $r_{n}=1$. Equations (\ref{en}) and (\ref{ep}) are written for the positive-frequency branches; the corresponding expressions for negative frequencies follow by exchanging the sublattice indices $1\leftrightarrow2$.

\begin{figure*}[htbp]
\includegraphics{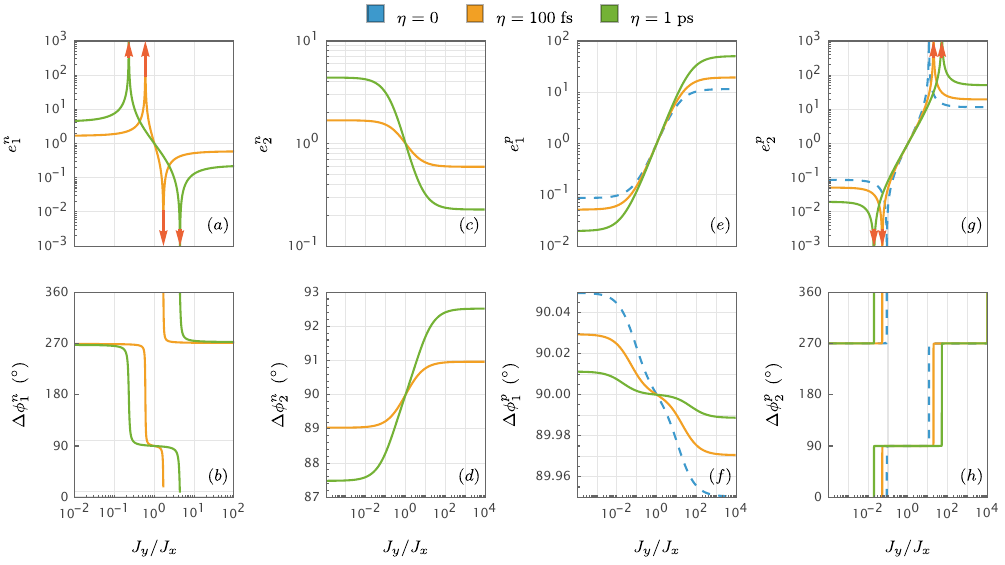}\caption{(color online). Ellipticity (upper row) and inter-component phase difference (lower row) as functions of the current ratio $J_{y}/J_{x}$ at positive frequencies for several values of~$\eta$. Panels (a)-(d) present the nutational modes of sublattices~1 and~2, while (e)-(h) display the corresponding precessional modes. The curves are calculated with Eqs. (\ref{en}), (\ref{ep}), and (\ref{phase dfferences at resonance}). Parameters are identical to those in Fig. \ref{spectrum_all}.}
\label{ellipticity_and_phase_difference}
\end{figure*}

The degree of polarization, characterized by the ellipticity, strongly depends on the ratio $J_{y}/J_{x}$ and the inertial relaxation time $\eta$, as shown in the upper panels of Fig. \ref{ellipticity_and_phase_difference}. Given similar behaviors for opposite signs of $J_{x,y}$ and resonant frequency, the analysis is restricted to $J_{x,y}>0$ and positive frequencies.

For the nutational mode, $\mathbf{m}_{1}$ becomes linearly polarized along the $x$ axis ($y$ axis) when $J_{y}/J_{x}=r_{n}$ ($J_{y}/J_{x}=1/r_{n}$), as indicated by the vanishing (diverging) ellipticity $e_{1}^{n}$ in Fig. \ref{ellipticity_and_phase_difference}(a) with red arrows. In contrast, Fig. \ref{ellipticity_and_phase_difference}(c) shows that $e_{2}^{n}$ remains finite and nonunity at these points, indicating elliptical polarization for $\mathbf{m}_{2}$. Thus, the nutational mode can be linearly polarized in one sublattice and elliptically polarized in the other.

A similar feature appears in the precessional mode: At $J_{y}/J_{x}=r_{p}$ ($J_{y}/J_{x}=1/r_{p}$), $\mathbf{m}_{2}$ becomes linearly polarized along the $x$-axis ($y$-axis), while $\mathbf{m}_{1}$ remains elliptically polarized, as seen in Figs. \ref{ellipticity_and_phase_difference}(e) and \ref{ellipticity_and_phase_difference}(g). The linear polarization is indicated with red arrows.

\begin{figure*}[htbp]
\selectlanguage{american}%
\includegraphics{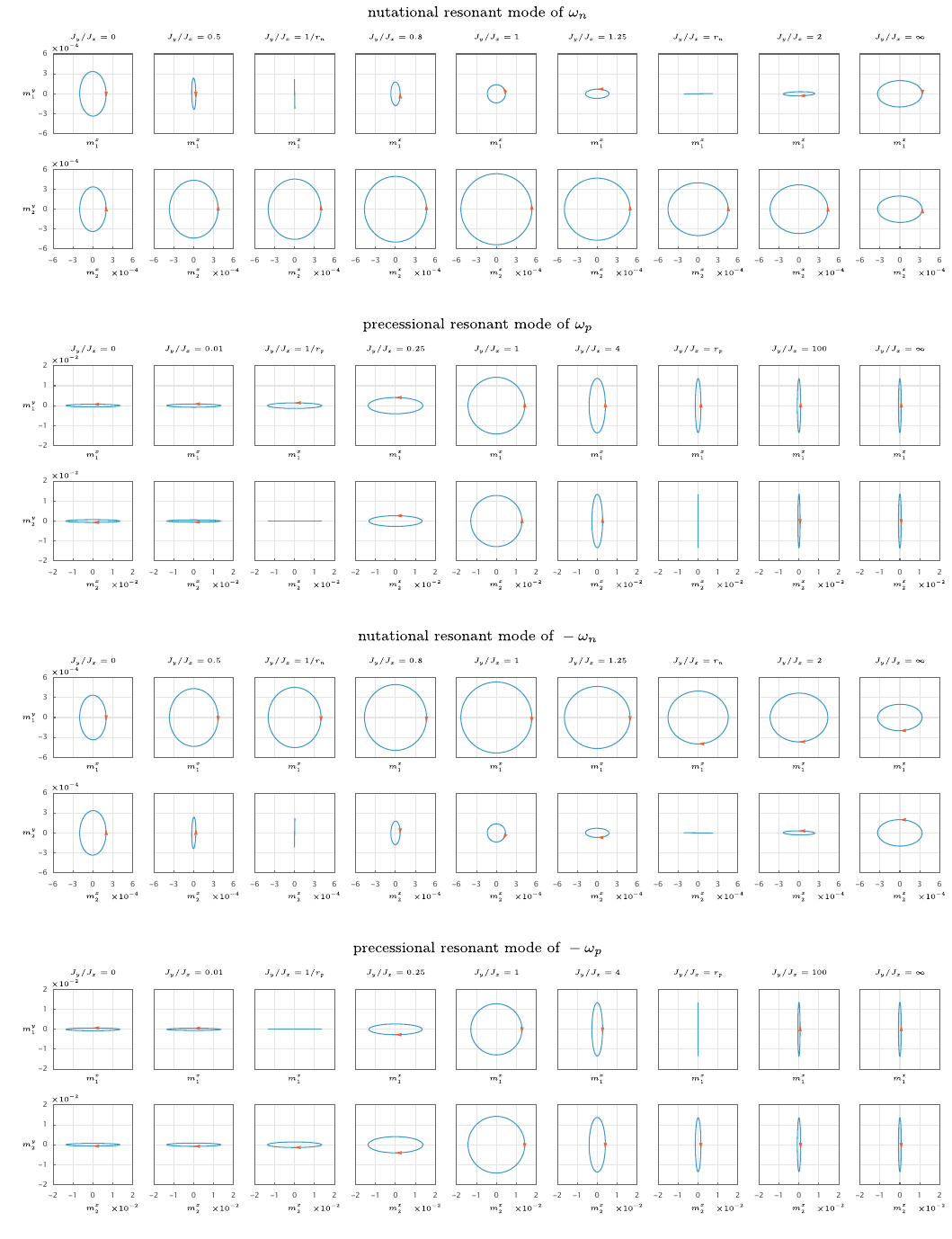}

\selectlanguage{english}%
\caption{(color online). Polarization and handedness of resonant modes for four branches ($\pm\omega_{p}$, $\pm\omega_{n}$) at various values of $J_{y}/J_{x}$. The inertial relaxation time is set to $\eta=100$ fs, yielding $r_{n}=1.6855$ and $r_{p}=19.4488$ from Eq. (\ref{rnp}). In each panel, the horizontal and vertical axes correspond to $m_{x}$ and $m_{y}$ components, respectively. Curves represent the magnetization orbits, and arrows indicate the sense of rotation. In the panels in columns 1-5, $J_{x}=1$GA$/$m$^{2}$ and $J_{y}$ is calculated by the ratio $J_{y}/J_{x}$. In the panels in columns 6-9, $J_{y}=1$GA$/$m$^{2}$ and $J_{x}$ is calculated by the ratio.}
\label{orbit}
\end{figure*}

In the limiting case where $J_{x}$ or $J_{y}$ vanishes, all modes are elliptically polarized with ellipticity equal to $r_{n,p}$ or $1/r_{n,p}$. When $J_{x}=J_{y}$, the ellipticity becomes unity, and all modes are circularly polarized. These polarization characteristics are also illustrated schematically in Fig. \ref{orbit} for representative values of $J_{y}/J_{x}$, in agreement with Fig. \ref{ellipticity_and_phase_difference}. Comparison of the ellipticity curves for different $\eta$ values reveals that increasing spin inertia enhances the deviation of $e_{1,2}^{n}$ and $e_{1,2}^{p}$ from unity, thereby increasing the ellipticity and, thus, the degree of polarization.

\begin{figure*}[t]
\selectlanguage{american}%
\includegraphics{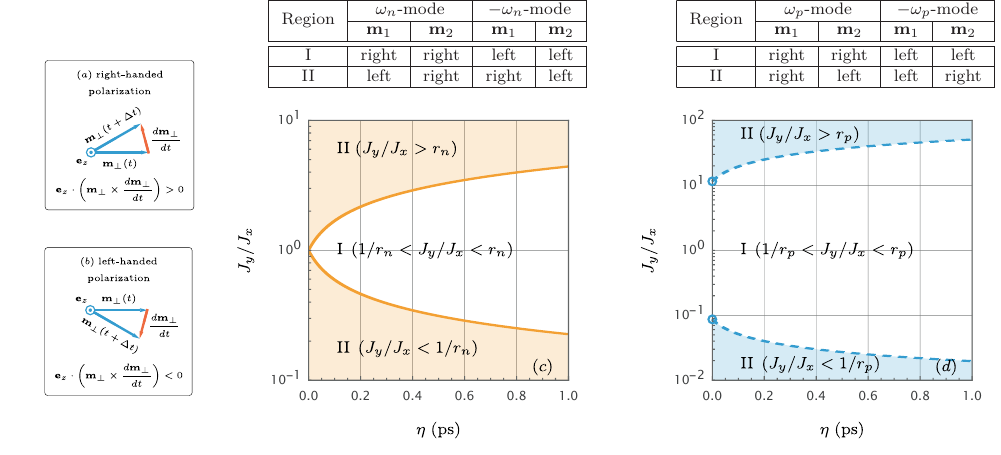}

\selectlanguage{english}%
\caption{(color online) (a) and (b) Definition of handedness. Phase diagrams illustrating the polarization of $\mathbf{m}_{1,2}$ as functions of $J_{y}/J_{x}$ and $\eta$ for (c) nutational and (d) precessional resonances. In (c) the upper (lower) boundary is given by $J_{y}/J_{x}=r_{n}$ ($J_{y}/J_{x}=1/r_{n}$); in (d) by $J_{y}/J_{x}=r_{p}$ ($J_{y}/J_{x}=1/r_{p}$). The upper and lower circles in (d) mark $J_{y}/J_{x}=\sqrt{(2\omega_{E}+\omega_{K})/\omega_{K}}$ and $\sqrt{\omega_{K}/(2\omega_{E}+\omega_{K})}$, respectively. The tables above (c) and (d) contain the handedness in different regions for nutational and precessional resonant modes, respectively.}
\label{polarization}
\end{figure*}

In addition to ellipticity, the handedness -- the sense of rotation of $\mathbf{m}_{1,2}$ around the $z$ axis -- completes the description of the polarization. Let $\mathbf{m}_{\perp}=m_{k}^{x}\mathbf{e}_{x}+m_{k}^{y}\mathbf{e}_{y}$ denote the in-plane magnetization. As outlined in Figs. \ref{polarization}a and \ref{polarization}b, a mode is righthanded if $\mathbf{e}_{z}\cdot(\mathbf{m}_{\perp}\times d\mathbf{m}_{\perp}/dt)>0$ or, equivalently, if $\omega\sin(\phi_{k}^{x}-\phi_{k}^{y})>0$ according to Eq. (\ref{spin waves}). The sign inversion corresponds to the opposite handedness.

At resonance, the phase differences defined in Eq. (\ref{phase dfferences}) can be simplified as 
\begin{equation}
\Delta\phi_{1,2}^{n(p)}=\tan^{-1}\left[X_{1,2}^{n(p)},Y_{1,2}^{n(p)}\right],\label{phase dfferences at resonance}
\end{equation}
where, for the $\omega_{n}$($\omega_{p}$)-branch resonance, 
\begin{eqnarray}
X_{1,2}^{n(p)} & = & \pm\alpha\omega_{n(p)}^{2}\left(J_{y}^{2}-J_{x}^{2}\right),\label{X}
\end{eqnarray}
\begin{eqnarray}
Y_{1,2}^{n}\! & = & \!\mp\frac{\kappa_{n}\!-\!2\eta\omega_{E}}{4\eta^{2}}\left(\sqrt{\kappa_{n}\!-\!2\eta\omega_{E}}J_{x}\!\mp\!\sqrt{\kappa_{n}\!+\!2\eta\omega_{E}}J_{y}\right)\nonumber \\
 &  & \times\left(\sqrt{\kappa_{n}\!+\!2\eta\omega_{E}}J_{x}\!\mp\!\sqrt{\kappa_{n}\!-\!2\eta\omega_{E}}J_{y}\right),\\
Y_{1,2}^{p}\! & = & \!\pm\frac{\kappa_{p}\!+\!2\eta\omega_{E}}{4\eta^{2}}\left(\sqrt{\kappa_{p}\!-\!2\eta\omega_{E}}J_{x}\!\pm\!\sqrt{\kappa_{p}\!+\!2\eta\omega_{E}}J_{y}\right)\nonumber \\
 &  & \times\left(\sqrt{\kappa_{p}\!+\!2\eta\omega_{E}}J_{x}\!\pm\!\sqrt{\kappa_{p}\!-\!2\eta\omega_{E}}J_{y}\right).\label{Yp}
\end{eqnarray}
where $\kappa_{n,p}$ has been defined in Eq. (\ref{kappa}). The phase differences for the $-\omega_{n,p}$ branches follow by exchanging the sublattice indices $1\leftrightarrow2$ in Eqs. (\ref{X})-(\ref{Yp}). In deriving Eq. (\ref{phase dfferences at resonance}) we insert the resonant frequencies for $\omega$ and use the approximations $\alpha^{2}\ll1$, $\alpha\beta\ll1$, and $\beta^{2}\ll1$. For positive resonant frequencies (righthanded SOTs) the mode is righthanded when $0^{\circ}<\Delta\phi_{1,2}^{n(p)}<180^{\circ}$ and lefthanded when $180^{\circ}<\Delta\phi_{1,2}^{n(p)}<360^{\circ}$.

Analogous to the ellipticity, the phase difference $\Delta\phi_{1,2}^{n(p)}$ between the $x$- and $y$-components is governed by the ratio $J_{y}/J_{x}$ (lower panels of Fig.~\ref{ellipticity_and_phase_difference}). For the nutational branch, the phase difference is $\Delta\phi_{1}^{n}=0^{\circ}$ at $J_{y}/J_{x}=r_{n}$ and $\Delta\phi_{1}^{n}=180^{\circ}$ at $J_{y}/J_{x}=1/r_{n}$ {[}Fig.~\ref{ellipticity_and_phase_difference}(b){]}. Hence $m_{1}^{x}$ and $m_{1}^{y}$ oscillate in phase at $J_{y}/J_{x}=r_{n}$ or antiphase at $J_{y}/J_{x}=1/r_{n}$, whereas $\mathbf{m}_{1}$ is linearly polarised. In the precessional branch, $\Delta\phi_{2}^{p}=180^{\circ}$ ($\Delta\phi_{2}^{p}=0^{\circ}$) at $J_{y}/J_{x}=r_{p}$ ($J_{y}/J_{x}=1/r_{p}$) {[}Fig. \ref{ellipticity_and_phase_difference}(h){]}, again indicating that $\mathbf{m}_{2}$ is linear polarized. The nutational mode $\mathbf{m}_{2}$ and precessional mode $\mathbf{m}_{1}$ exhibit the opposite trends with increasing the inertial relaxation time $\eta$, i.e. the phase difference $\Delta\phi_{2}^{n}$ increases with $\eta,$whereas $\Delta\phi_{1}^{p}$ decreases.

In the limiting case $J_{x}\to0$ or $J_{y}\to0$, and using $\omega_{K}\ll\omega_{E}$, in Eqs. (\ref{X})--(\ref{Yp}), one finds $\Delta\phi\to270^{\circ}$ for the nutational mode of $\mathbf{m}_{1}$ and the precessional mode of $\mathbf{m}_{2}$, indicating lefthanded polarization. In contrast, $\Delta\phi$ approaches $90^{\circ}$ for the nutational mode of $\mathbf{m}_{2}$ and the precessional mode of $\mathbf{m}_{1}$, corresponding to righthanded polarization. When $J_{x}=J_{y}$, all modes exhibit $\Delta\phi=90^{\circ}$, indicating right-handed circular polarization.

Using Eqs. (\ref{phase dfferences at resonance})-(\ref{Yp}), the polarization handedness of the nutational and precessional resonant modes in sublattices 1 and 2 can be determined by the sign of $\omega_{r}\sin\Delta\phi_{1,2}^{n(p)}$, where $\omega_{r}$ denotes the resonant frequency. According to this criterion, the condition $\sin\Delta\phi_{1,2}^{n(p)}=0$ defines the boundaries between regions of different handedness. These boundaries correspond precisely to $J_{y}/J_{x}=r_{n,p}$ and $J_{y}/J_{x}=1/r_{n,p}$, where $r_{n,p}$ is given in Eq. (\ref{rnp}).

Based on this analysis, we construct the phase diagrams shown in Figs. \ref{polarization}(c) and \ref{polarization}(d), mapping the polarization types in the parameter space spanned by $\eta$ and $J_{y}/J_{x}$. The boundary curves in Fig. \ref{polarization} are calculated using the approximate analytical expression (\ref{rnp}). The shaded and unshaded regions are determined by evaluating the sign of $\omega_{r}\sin\Delta\phi_{1,2}$, where $\Delta\phi_{1,2}$ is computed from Eq. (\ref{phase dfferences}) using the corresponding resonant frequencies. The analytical approximations agree very well with the exact numerical results.

From the above analysis -- particularly Figs. \ref{orbit} and \ref{polarization} -- it is evident that the polarization of resonant modes can be tuned by controlling both the handedness (i.e., the sign of the frequency) and the ellipticity ($J_{y}/J_{x}$) of the alternating driving force (here, SOTs generated by alternating currents). For the sake of clarity, we restrict the discussion to positive $J_{x,y}$, such that the handedness of the driving force is solely determined by the sign of its frequency.

For both nutational and precessional modes, the polarization of all resonant modes matches that of the SOTs when the ellipticity of the drive is close to $1$ that is, when $1/r_{n}<J_{y}/J_{x}<r_{n}$ (nutation) or $1/r_{p}<J_{y}/J_{x}<r_{p}$ (precession). This corresponds to Region I in Figs.~\ref{polarization}(c) and \ref{polarization}(d). In this regime, both $\mathbf{m}_{1}$ and $\mathbf{m}_{2}$ rotate righthandedly about the $+z$ axis for positive resonant frequencies, and lefthandedly for negative frequencies, consistent with the behavior shown schematically in Fig. \ref{orbit}.

When the ellipticity of the SOTs deviates more strongly from $1$, the polarization behavior becomes sublattice dependent {[}Region II in Figs. \ref{polarization}(c) and \ref{polarization}(d){]}. Specifically, for nutational modes in Region II ($J_{y}/J_{x}<1/r_{n}$ or $J_{y}/J_{x}>r_{n}$), $\mathbf{m}_{1}$ rotates lefthandedly while $\mathbf{m}_{2}$ maintains righthanded rotation at positive frequency. For precessional modes in Region II ($J_{y}/J_{x}<1/r_{p}$ or $J_{y}/J_{x}>r_{p}$), the handedness is reversed: $\mathbf{m}_{1}$ rotates righthandedly, while $\mathbf{m}_{2}$ rotates lefthandedly for positive frequency. We emphasize that the boundaries of the polarization regions depend sensitively on the inertial relaxation time $\eta$, enabling its experimental determination. By tuning the ratio of two orthogonally applied currents and measuring the resulting mode polarizations, $\eta$ can be extracted by fitting the data with Eqs.~(\foreignlanguage{american}{\ref{en}-\ref{rnp}).}

\selectlanguage{american}%
From Figs. \foreignlanguage{english}{\ref{polarization}(c) and \ref{polarization}(d)} we infer that Region~I of the nutational mode is much smaller than that of the precessional mode because $r_{n}\ll r_{p}$. This is consistent with Eq.~\eqref{rnp}, which for $\omega_{K}/\omega_{E}\ll1$ gives 
\begin{equation}
\begin{aligned}r_{n} & =\sqrt{1+2\eta\omega_{E}}+O\left[\omega_{K}\right],\\
r_{p} & =\sqrt{\frac{2\omega_{E}}{\omega_{K}}}\sqrt{1+2\eta\omega_{E}}+O\left[\omega_{K}^{1/2}\right].
\end{aligned}
\end{equation}
For typical AFM parameters with $\omega_{K}\ll\omega_{E}$, it follows that $r_{n}\ll r_{p}$. Consequently, the handedness of the nutational mode reverses at a much smaller current difference than that of the precessional mode.

\selectlanguage{english}%
In FMs, as shown in Appendix \ref{resonant modes} and Fig. \ref{orbit_FM}, the handedness of resonant modes is independent of the ratio $J_{y}/J_{x}$. For positive (negative) frequency, the nutational mode is lefthanded (righthanded), while the precessional mode is righthanded (lefthanded). In contrast, the situation in AFMs is more intricate. As $J_{y}/J_{x}$ crosses the critical values $r_{n(p)}$ or $1/r_{n(p)}$, the handedness of the nutational (precessional) mode is reversed for one of the sublattices. These critical points depend on the inertial relaxation time $\eta$ and are a unique feature of AFMs, absent in the FM case.

Finally, the fieldlike SOT {[}the $\beta$ term in Eq. (\ref{SOTs}){]} has a negligible impact on the resonance modes. In the peak-height and phase-difference expressions, $\beta$ enters only through the factors $1+\alpha\beta$ and $1+\beta^{2}$, and the contribution of the $\beta$ term is small due to $\alpha\beta\ll1$ and $\beta^{2}\ll1$ for typical torques induced by the spin Hall effect \citep{AManchon}. Numerical simulations confirm the accuracy of the analytical approximation.

\section{Discussions}

First, it is worthwhile to elucidate the significance of the negative frequency. When solving the secular equation for a linear oscillating system, the eigenfrequencies usually come in a pair of opposite numbers. For natural oscillation of magnetization, there is no difference between the positive and negative frequency branches, apart from the opposite handedness of oscillating mode. Thus, the negative eigenfrequency has often been thought to be unnecessary. However, the case is somewhat complicated for resonance of magnetization. As shown in Figs. \ref{spectrum_all} and \ref{spectrum_all_FM}, the peak heights are different at positive and negative resonant frequencies. Also, Fig. \ref{orbit} indicates that the polarization and handedness of resonant modes are different for positive and negative resonant frequencies. The reason is that the frequency of forced oscillation belongs to the driving force. Different signs of frequency correspond to different handednesses of driving force. As illustrated in above sections, the properties of resonant mode depend on the handedness of driving force, resulting in the asymmetry between positive and negative resonant frequencies.

Second, AFMs host both lefthanded and righthanded spin waves \citep{SMRezende}, enabling storing information on the handedness (charity) of polarization. This feature has been proposed to design transistors \citep{RCheng,TYu} and logic gates \citep{WYu,CJia} based on spin waves. Thus, it is crucial to detect and manipulate the handedness of spin waves, which has been achieved by inverse spin Hall effect in synthetic AFMs \citep{YShiota}. With regard to the spin waves with zero wave vector, it has been shown that the polarization of resonant mode can be identified and manipulated by using different types of microwave magnetic fields \citep{XChen}. Here, by steering the degree of polarization ($J_{y}/J_{x}$) and the handedness of the alternating SOTs, we realize the switching of polarization and the reversal of handedness for the nutational and precessional resonant AFM modes. The critical value of $J_{y}/J_{x}$ can be tuned by the inertial relaxation time $\eta$, as indicated by Eq. (\ref{rnp}) and Fig. \ref{polarization}. On the other hand, by varying $J_{y}/J_{x}$ and observing the abrupt change of the polarization, one can measure the critical values $J_{y}/J_{x}=r_{n,p}$, and then estimate $\eta$, which, so far, is not experimentally determined for AFMs.

Third, it should be pointed out that we only explore the linear dynamics of AFMs driven by alternating SOTs. The results are suitable for weak driving forces. With the drives becoming stronger, some high-order harmonic magnetic responses gradually emerge. Thus, beyond the linear response described in Eq. (\ref{linear modes}), $m_{k}^{x,y}$ should be expanded into a series of harmonics, including the higher-order terms of $j_{x,y}$. Inserting this ansatz and its time derivative into the ILLG equations, and collecting terms at each harmonic ($e^{-ni\omega t},n=1,2,3,\cdots$), one arrives at the scalar equations about higher-order susceptibilities by matching the coefficients for every harmonic orders. In general, it is complicated to get explicit analytic solutions for the susceptibilities and some numeric treatments must be resorted to. Moreover, we use a minimal model of AFMs, in which a perpendicular uniaxial anisotropy is considered. For more complex geometries -- for example, involving the biaxial anisotropy, the Dzyaloshinskii-Moriya interaction, triangular spin textures -- the linearization technique can also be used. For noncollinear ground states (e.g., canted AFMs \citep{IBoventer}), fluctuations must first be defined in local frames attached to each sublattice, and the equations of motion are then linearized about those frames. Also, by use of the eigenvalues and eigenvectors of linearized system, one can explore the dispersion and polarization of inertial resonant modes. A more thorough understanding of the higher harmonics and further exploration on the complex AFMs are beyond the scope of this work and will be pursued in the future.

Finally, it is interesting to compare the roles of SOTs and spin-transfer torques in driving magnetic resonances. In SOT-based devices, two mutually perpendicular in-plane charge currents can be applied and tuned independently, generating two independent ac driving torques, analogous to two orthogonal in-plane ac magnetic fields. By adjusting the magnitudes and directions (and, if applicable, phases) of these currents, the total driving torque can be engineered to realize different polarizations and handedness. In contrast, in spin-transfer-torque structures the perpendicular current provides essentially a single driving torque with a fixed direction, which does not allow independent control of the polarization and handedness of the excitation.

\section{Conclusion}

\label{Conclusion}

In this paper, we investigate resonant magnetization precession and nutation in a bilayer system composed of a uniaxial AFM and an adjacent HM layer, driven by two in-plane, mutually perpendicular alternating currents. We find that, contrary to the common view that the handedness of a resonant mode is essentially fixed by the material, the polarization state and handedness of both precessional and nutational AFM modes can be controlled by the alternating SOTs generated by in-plane $J_{x}$ and $J_{y}$ currents. As the amplitude ratio $J_{y}/J_{x}$ of orthogonal driving currents is varied, AFM resonant modes evolve continuously from elliptical to circular and, at critical ratios $r_{n,p}$, to linear polarization. We derive the critical ratios $r_{n,p}$ for the nutational ($n$) and precessional ($p$) modes at which a mode becomes linearly polarized and the mode handedness reverse. Moreover, a given mode can be linear on one AFM sublattice while remaining elliptical on the other.

We find that the critical ratios depend explicitly on the inertial relaxation time $\eta$ and differ between the nutational and precessional resonances, enabling inertia-controlled chirality switching. The polarization and handedness reversals at $r_{n,p}$ provide a practical method to determine inertial relaxation time in SOT-driven AFM resonance experiments.

This tunability is not available in the FM case, where the resonant modes remain nearly circular. A detailed comparison with the FM ILLG case shows that in FMs the nutational mode is always lefthanded and the precessional mode always righthanded, and both remain nearly circular regardless of the drive ellipticity. In AFMs, by contrast, both ellipticity and handedness of the modes depend sensitively on $J_{y}/J_{x}$, $\eta$, and the sign of $\omega$. Both nutational and precessional modes appear in pairs at $\pm\omega_{n,p}$ unlike the FM case where certain branches are strongly suppressed for circular driving.

Our findings highlight the richer landscape of polarization and handedness in AFM resonant dynamics and suggest potential applications for AFM modes as carriers of information in spintronic devices. This work advances the fundamental understanding of inertial resonant modes and provides a basis for chirality control in spin-wave-based information processing. We expect that the control of the nutational-mode handedness permits a robust manipulation of nutation magnons in AFMs.

\section{Acknowledgments}

Peng-Bin He was supported by the NSF of Changsha City (Grant No. kq2208008) and the NSF of Hunan Province (Grant No. 2023JJ30116). Ri-Xing Wang was supported by the key Program of Education Bureau of Hunan Province (Grant No.24A0494) and the Regional Joint Funds of the NSF of Hunan Province (Grant No.2024JJ7312).

\appendix
%dummy comment inserted by tex2lyx to ensure that this paragraph is not empty

\section{Linearization of AFM system}

\label{app A}

To linearize the coupled two-lattices AFM system, it is convenient to expand Eq. (\ref{LLG equation}) in the component form
\begin{widetext}
\begin{eqnarray}
\frac{dm_{k}^{x}}{dt} & = & \omega_{E}\left(m_{k}^{y}m_{3-k}^{z}-m_{k}^{z}m_{3-k}^{y}\right)-\omega_{K}m_{k}^{y}m_{k}^{z}+\alpha\left(m_{k}^{y}\frac{dm_{k}^{z}}{dt}-m_{k}^{z}\frac{dm_{k}^{y}}{dt}\right)+\eta\left(m_{k}^{y}\frac{d^{2}m_{k}^{z}}{dt^{2}}-m_{k}^{z}\frac{d^{2}m_{k}^{y}}{dt^{2}}\right)\nonumber \\
 &  & -\rho\left[j_{x}m_{k}^{x}m_{k}^{y}+j_{y}\left(m_{k}^{y2}+m_{k}^{z2}\right)\right]+\rho\beta j_{x}m_{k}^{z},\label{A1}\\
\frac{dm_{k}^{y}}{dt} & = & \omega_{E}\left(m_{k}^{z}m_{3-k}^{x}-m_{k}^{x}m_{3-k}^{z}\right)-\omega_{K}m_{k}^{z}m_{k}^{x}+\alpha\left(m_{k}^{z}\frac{dm_{k}^{x}}{dt}-m_{k}^{x}\frac{dm_{k}^{z}}{dt}\right)+\eta\left(m_{k}^{z}\frac{d^{2}m_{k}^{x}}{dt^{2}}-m_{k}^{x}\frac{d^{2}m_{k}^{z}}{dt^{2}}\right)\nonumber \\
 &  & +\rho\left[j_{y}m_{k}^{x}m_{k}^{y}+j_{x}\left(m_{k}^{z2}+m_{k}^{x2}\right)\right]+\rho\beta j_{y}m_{k}^{z},\\
\frac{dm_{k}^{z}}{dt} & = & \omega_{E}\left(m_{k}^{x}m_{3-k}^{y}-m_{k}^{y}m_{3-k}^{x}\right)+\alpha\left(m_{k}^{x}\frac{dm_{k}^{y}}{dt}-m_{k}^{y}\frac{dm_{k}^{x}}{dt}\right)+\eta\left(m_{k}^{z}\frac{d^{2}m_{k}^{y}}{dt^{2}}-m_{k}^{y}\frac{d^{2}m_{k}^{x}}{dt^{2}}\right)\nonumber \\
 &  & -\rho\left(j_{x}m_{k}^{y}m_{k}^{z}-j_{y}m_{k}^{z}m_{k}^{x}\right)-\rho\beta\left(j_{x}m_{k}^{x}+j_{y}m_{k}^{y}\right),\label{A3}
\end{eqnarray}
where $k=1,2$, denoting the two sublattices. In the main text, it has been assumed that the static equilibrium magnetization lies along the $z$ direction. Thus, for small deviations ($m_{k}^{x}$ and $m_{k}^{y}$) from equilibrium, $m_{k}^{z}$ is unchanged to first-order in small quantities ($m_{1}^{z}\approx1$ and $m_{2}^{z}\approx-1$). For resonance, the excitation sources $j_{x,y}$ are also small quantities. Then, keeping only terms linear in $m_{k}^{x}$, $m_{k}^{y}$, $j_{x}$, and $j_{y}$ in Eqs. (\ref{A1})-(\ref{A3}), one can obtain the linear differential equations about $m_{1}^{x}$, $m_{1}^{y}$, $m_{2}^{x}$, and $m_{2}^{y}$, expressed in the matrix form as %\begin{small}

\begin{equation}
\left(\begin{array}{cccc}
1 & \alpha & 0 & 0\\
-\alpha & 1 & 0 & 0\\
0 & 0 & 1 & -\alpha\\
0 & 0 & \alpha & 1
\end{array}\right)\frac{d}{dt}\left(\begin{array}{c}
m_{1}^{x}\\
m_{1}^{y}\\
m_{2}^{x}\\
m_{2}^{y}
\end{array}\right)=\left(\begin{array}{cccc}
0 & -\omega_{E}-\omega_{K} & 0 & -\omega_{E}\\
\omega_{E}+\omega_{K} & 0 & \omega_{E} & 0\\
0 & \omega_{E} & 0 & \omega_{E}+\omega_{K}\\
-\omega_{E} & 0 & -\omega_{E}-\omega_{K} & 0
\end{array}\right)\left(\begin{array}{c}
m_{1}^{x}\\
m_{1}^{y}\\
m_{2}^{x}\\
m_{2}^{y}
\end{array}\right)+\rho\left(\begin{array}{cc}
\beta & -1\\
1 & \beta\\
-\beta & -1\\
1 & -\beta
\end{array}\right)\left(\begin{array}{c}
j_{x}\\
j_{y}
\end{array}\right).\label{linearized equation}
\end{equation}
%\end{small}
\end{widetext}

Given that the presumed solutions and the applied currents have a form $\sim e^{-i\omega t}$, Eq. (\ref{linear modes}) can be obtained from Eq. (\ref{linearized equation}) by computing the inverse of matrix.

\section{steady-state linear modes in FMs driven by SOTs}

The ILLG equation of FMs driven by SOTs reads 
\begin{equation}
\frac{d\mathbf{m}}{dt}=\mathbf{m}\times\frac{d\mathcal{E}}{d\mathbf{m}}+\alpha\mathbf{m}\times\frac{d\mathbf{m}}{dt}+\eta\mathbf{m}\times\frac{d^{2}\mathbf{m}}{dt^{2}}+\boldsymbol{\tau},\label{LLG equation of FM}
\end{equation}
where $\mathbf{m}$ is the unit vector of magnetization. The magnetic energy arises from perpendicular magnetocrystalline anisotropy and the demagnetizing field (within the local approximation), and in frequency units it is given by 
\begin{equation}
\mathcal{E}=\omega_{K}\left(\mathbf{m}\cdot\mathbf{e}_{z}\right)^{2},
\end{equation}
where $\omega_{K}=\gamma_{0}H_{K}$, with $H_{K}$ being effective anisotropy field defined by $H_{K}=2K_{u}/M_{s}-\mu_{0}M_{s}$, with $K_{u}$ being the magnetocrystalline anisotropy constant, $\mu_{0}$ the vacuum susceptibility, and $M_{s}$ the saturation magnetization. The SOTs $\mathbf{\boldsymbol{\tau}}$ read, 
\begin{equation}
\mathbf{\boldsymbol{\tau}}=-\rho\left\{ \mathbf{m}\times\left[\mathbf{m}\times\left(\mathbf{e}_{z}\times\mathbf{j}_{e}\right)\right]+\beta\mathbf{m}\times\left(\mathbf{e}_{z}\times\mathbf{j}_{e}\right)\right\} ,
\end{equation}
where $\rho$, $\beta$, and $\mathbf{j}_{e}$ are the same as those in Eq. (\ref{SOTs}).

By the same procedure as Sec. \ref{resonant spectrum}, the steady-state linear modes are derived as 
\begin{equation}
\left[\begin{array}{c}
m_{x}(t)\\
m_{y}(t)
\end{array}\right]=\chi\left[\begin{array}{c}
j_{x}(t)\\
j_{y}(t)
\end{array}\right]\label{linear modes of FM}
\end{equation}
In Eq. (\ref{linear modes of FM}), the susceptibility reads 
\begin{equation}
\chi=\frac{\rho}{\Delta}\left(\begin{array}{cc}
\chi_{1} & \chi_{2}\\
-\chi_{2} & \chi_{1}
\end{array}\right),
\end{equation}
where $\chi_{1}=\Omega+i(\alpha-\beta)\omega$, $\chi_{2}=\beta\Omega+i(1+\alpha\beta)\omega$, and 
\begin{equation}
\Delta=\omega^{2}+\Omega^{2}+i\alpha\omega,
\end{equation}
with $\Omega=\omega_{K}-\eta\omega^{2}$. Taking the real parts of Eq. (\ref{linear modes of FM}), the linear modes can be written explicitly as 
\begin{equation}
m_{x(y)}=\mathcal{A}_{x(y)}\cos\left[\omega t+\phi_{x(y)}\right].\label{spin wave of FM}
\end{equation}
In Eq. (\ref{spin wave of FM}), the amplitudes are expressed as 
\begin{eqnarray}
\mathcal{A}_{x(y)} & = & \rho\sqrt{\frac{P_{1}J_{x(y)}^{2}+P_{2}J_{y(x)}^{2}+P_{3}J_{x}J_{y}}{\widetilde{\Delta}}},\label{amplitude of FM}
\end{eqnarray}
where $P_{1}=(\alpha-\beta)^{2}\omega^{2}+\Omega^{2}$, $P_{2}=(1+\alpha\beta)^{2}\omega^{2}+\beta^{2}\Omega^{2}$, $P_{3}=2(1+\beta^{2})\Omega\omega$, and 
\begin{equation}
\widetilde{\Delta}=\left[\left(\Omega+\omega\right)^{2}+\alpha^{2}\omega^{2}\right]\left[\left(\Omega-\omega\right)^{2}+\alpha^{2}\omega^{2}\right].
\end{equation}

The phases in Eq. (\ref{spin wave of FM}) are given by 
\begin{eqnarray}
\phi_{x}\! & = & \!\tan^{-1}\left(Q_{1}J_{x}\!+\!Q_{2}J_{y},Q_{3}J_{x}\!+\!Q_{4}J_{y}\right),\label{phase of mx of FM}\\
\phi_{y}\! & = & \!\tan^{-1}\left(Q_{4}J_{x}\!+\!Q_{3}J_{y},-Q_{2}J_{x}\!-\!Q_{1}J_{y}\right),\label{phase of my of FM}
\end{eqnarray}
where $\tan^{-1}(x,y)$ denotes the two-argument arctangent, which correctly identifies the quadrant of the point $(x,y)$ \citep{arctan}. The frequency-dependent parameters $Q_{i}$ are defined as $Q_{1}=\Omega[(1+2\alpha\beta-\alpha^{2})\omega^{2}-\Omega^{2}]$, $Q_{2}=\omega[(1+\alpha^{2})(1+\alpha\beta)\omega^{2}-(1-\alpha\beta)\Omega^{2}]$, $Q_{3}=\omega[(1+\alpha^{2})(\alpha-\beta)\omega^{2}+(\alpha+\beta)\Omega^{2}]$, and $Q_{4}=\Omega[(2\alpha-\beta+\alpha^{2}\beta)\omega^{2}+\beta\Omega^{2}]$. From Eqs. (\ref{phase of mx of FM}) and (\ref{phase of my of FM}), the phase difference between $m_{x}$ and $m_{y}$ is calculated as 
\begin{equation}
\Delta\phi\equiv\phi_{x}-\phi_{y}=\tan^{-1}\left(X,Y\right),\label{phase dfferences of FM}
\end{equation}
where $X=[(\alpha-\beta)(1+\alpha\beta)\omega^{2}+\beta\Omega^{2}](J_{y}^{2}-J_{x}^{2})$, and $Y=(1+\beta^{2})\{\Omega\omega(J_{x}^{2}+J_{y}^{2})+[(1+\alpha^{2})\omega^{2}+\Omega^{2}]J_{x}J_{y}\}$.

\begin{figure}[t]
\selectlanguage{american}%
\includegraphics{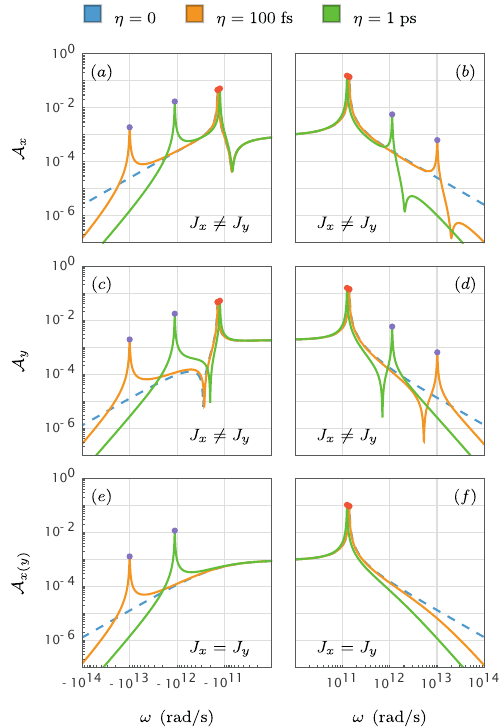}

\selectlanguage{english}%
\caption{(color online). FM resonant spectra for various values of $\eta$. (a)-(d) Under elliptically polarized SOTs ($J_{x}\protect\neq J_{y}$), both precessional and nutational FM resonances are excited on the positive- and negative-frequency ranges. (e) and (f) For circular polarization ($J_{x}=J_{y}$\LyXZeroWidthSpace ), the nutational mode is observed at negative frequencies, whereas the precessional mode appears at positive frequencies. Noninertial spectra at $\eta=0$ are shown with dashed curves. The precessional resonance is shifted by finite spin inertia ($\eta>0$) and additional nutational resonances are induced (solid curves). Peaks labeled with red (purple) points correspond to precessional (nutational) resonances. To enhance visibility, both axes are shown on a logarithmic scale, and low-frequency regions are omitted. All spectra are calculated from Eq. (\ref{amplitude of FM}) using the following parameters: $\omega_{K}=0.14$ THz, $\alpha=0.01$, $\beta=0.02$, $\rho\approx0.13$ Hz$/$(A$/$m$^{2}$); in (a)-(d), $J_{x}=1$ GA$/$m$^{2}$, $J_{y}=2$ GA$/$m$^{2}$; in (e) and (f), $J_{x}=J_{y}=1$ GA$/$m$^{2}$. For comparison with AFMs, the values of $\omega_{K}$ and $\rho$ are chosen to match those in Fig. \ref{spectrum_all}.}
\label{spectrum_all_FM}
\end{figure}

\section{resonant modes in FMs driven by SOTs}

\label{resonant modes}

Assuming $\alpha^{2}\ll1$, Eq. (\ref{amplitude of FM}) implies that the resonant frequency satisfies 
\begin{equation}
\eta\omega^{2}\pm\omega-\omega_{K}=0.
\end{equation}
Without spin inertia ($\eta=0$), the resonant frequency is $\pm\omega_{K}$. In the presence of spin inertia, the resonant frequencies of nutation $\pm\omega_{n}$ and precession $\pm\omega_{p}$ are given by 
\begin{eqnarray}
\omega_{n,p}=\frac{\sqrt{1+2\eta\omega_{K}}\pm1}{2\eta}.
\end{eqnarray}

From Eq. (\ref{amplitude of FM}), the heights of the resonant peaks can be derived. Under the approximations $\alpha^{2}\ll1$, $\alpha\beta\ll1$, and $\beta^{2}\ll1$, the peak heights are approximately equal for the $x$- and $y$-components of magnetization. Corresponding to the nutation and precession, the peak heights are written as 
\begin{eqnarray}
\mathcal{H}_{n}^{\pm} & = & \rho\frac{\eta}{\alpha}\frac{\left\vert J_{x}\mp J_{y}\right\vert }{\sqrt{1+2\eta\omega_{K}}+1},\label{peak height of nutation of FM}\\
\mathcal{H}_{p}^{\pm} & = & \rho\frac{\eta}{\alpha}\frac{\left\vert J_{x}\pm J_{y}\right\vert }{\sqrt{1+2\eta\omega_{K}}-1}.\label{peak height of precession of FM}
\end{eqnarray}
Adopting the same approximation and substituting $\omega$ into Eq. (\ref{phase dfferences of FM}), the phase differences corresponding to the $\pm\omega_{n}$ and $\pm\omega_{p}$ resonant modes are given by 
\begin{eqnarray}
\Delta\phi_{n}^{\pm} & = & \tan^{-1}\left[\alpha\left(J_{y}^{2}-J_{x}^{2}\right),\mp\left(J_{x}\mp J_{y}\right)^{2}\right],\label{phase difference of nutation of FM}\\
\Delta\phi_{p}^{\pm} & = & \tan^{-1}\left[\alpha\left(J_{y}^{2}-J_{x}^{2}\right),\pm\left(J_{x}\pm J_{y}\right)^{2}\right].\label{phase difference of precession of FM}
\end{eqnarray}

\begin{figure}[H]
\selectlanguage{american}%
\includegraphics{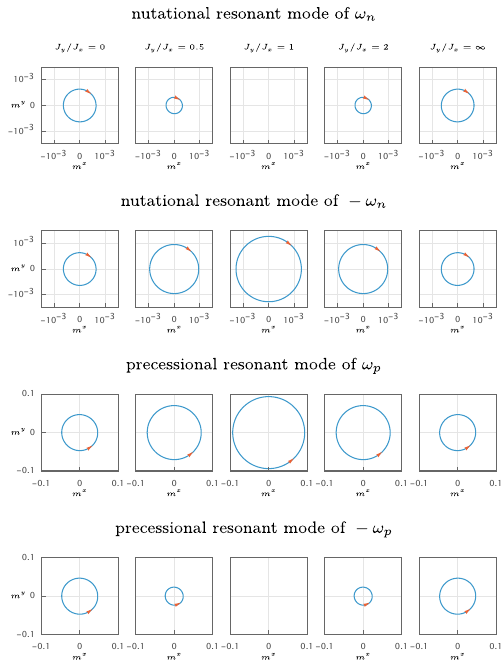}

\selectlanguage{english}%
\caption{(color online). Polarization and handedness of resonant modes for various ratios $J_{y}/J_{x}$. The inertial relaxation time is set to $\eta=100$ fs. In each panel, the horizontal and vertical axes represent the $m_{x}$ and $m_{y}$ components, respectively. The curves depict the trajectories of the magnetization vector, and the arrows indicate the sense of rotation. In the panels in columns 1-3, $J_{x}=1$GA$/$m$^{2}$ and $J_{y}$ is calculated by the ratio $J_{y}/J_{x}$. In the panels in columns 4-5, $J_{y}=1$GA$/$m$^{2}$ and $J_{x}$ is calculated by the ratio.}
\label{orbit_FM}
\end{figure}

Eqs. (\ref{peak height of nutation of FM}) and (\ref{peak height of precession of FM}) imply that, for the circularly polarized SOTs ($J_{x}=J_{y}$), the nutational mode at $\omega_{n}$ and the precessional mode at $-\omega_{p}$ are nearly suppressed, as shown in Figs.~\ref{spectrum_all_FM}(e) and~\ref{spectrum_all_FM}(f). Meanwhile, Eqs. (\ref{phase difference of nutation of FM}) and (\ref{phase difference of precession of FM}) indicate that the phase differences corresponding to the $-\omega_{n}$ and $\omega_{p}$ resonances are both $90^{\circ}$. Consequently, one finds $-\omega_{n}\sin\Delta\phi_{n}^{-}<0$ and $\omega_{p}\sin\Delta\phi_{p}^{+}>0$, indicating that only the lefthanded nutational mode and righthanded precessional mode are excited by circularly polarized SOTs. This is also illustrated in the third column of Fig. \ref{orbit_FM}. These results qualitatively agree with Ref.~\citep{RMondal_PRB104_21}, where a circularly polarized oscillating magnetic field was used to drive the system.

However, for elliptically polarized SOTs ($J_{x}\neq J_{y}$), all resonant modes -- both at positive and negative frequencies -- can be excited, as shown in Figs. \ref{spectrum_all_FM}(a)-(d). Moreover, all modes are nearly circularly polarized. For nutational resonances, the phase differences satisfy $180^{\circ}<\Delta\phi_{n}^{+}<360^{\circ}$ and $0^{\circ}<\Delta\phi_{n}^{-}<180^{\circ}$, leading to $\pm\omega_{n}\sin\Delta\phi_{n}^{\pm}<0$. Therefore, both the positive- and negative-frequency nutational modes (first and third rows of Fig. \ref{orbit_FM}, respectively) are lefthanded. Similarly, for precessional resonances, one has $0^{\circ}<\Delta\phi_{p}^{+}<180^{\circ}$ and $180^{\circ}<\Delta\phi_{p}^{-}<360^{\circ}$, which results in $\pm\omega_{p}\sin\Delta\phi_{p}^{\pm}>0$. Consequently, both the positive- and negative-frequency precessional modes (second and fourth rows of Fig. \ref{orbit_FM}) are righthanded. In summary, regardless of whether the SOT is righthanded or lefthanded, nutational resonant modes in FMs are always lefthanded, while precessional resonant modes are always righthanded.

\section*{References}

\bibliographystyle{apsrev4-2-titles}
\bibliography{converted_bibtex_1}
\selectlanguage{american}%

\end{document}